\documentclass[11pt]{article}
\usepackage{latexsym}  
\usepackage{amssymb}
\usepackage{graphicx}
\usepackage{amsmath}

\begin{document}

\begin{center}

{\Large\bf Empirical Neutrino Mass Matrix 
Related to Up-Quark Masses}

\vspace{3mm}
{\bf Yoshio Koide}

{\it IHERP, Osaka University, 1-16 Machikaneyama, 
Toyonaka, Osaka 560-0043, Japan} \\
{\it E-mail address: koide@het.phys.sci.osaka-u.ac.jp}

\end{center}

\vspace{3mm}
\begin{abstract}
Based on an approach to quark and lepton masses, where
the mass spectra originate in vacuum expectation values
of U(3) flavor nonet (gauge singlet) scalars, neutrino 
masses and mixing are investigated.  As an offshoot of 
this approach, it is found that an empirical neutrino mass 
matrix which is described in terms of the up-quark 
and charged lepton masses can accommodate to a nearly 
tribimaximal mixing. 
\end{abstract}

\vspace{3mm}
\noindent{\large\bf 1 \ Introduction}

The observed neutrino oscillation data are in favor of 
the so-called ``tribimaximal mixing" \cite{tribi}.
Usually, the mixing matrix form  has been understood 
based on a discrete symmetry.
One of the motivation of the present paper is to 
investigate whether such a nearly tribimaximal mixing
can be understood without assuming such a discrete symmetry. 
We will find an empirical neutrino mass matrix which is
described in terms of up-quark and charged lepton mass 
matrices and which leads to a nearly tribimaximal mixing
without assuming any discrete symmetry.

In this paper, we will discuss a non-standard approach to the 
masses and mixings of the quarks and leptons against the
conventional mass matrix models. 
In this model, the mass spectra of the quarks
and leptons originates in a vacuum expectation value
(VEV) structure of a U(3)-flavor nonet
scalar $\Phi$ \cite{Koide90,YK-Tanimoto96,YK-JHEP07}.
In the present approach, we write a superpotential $W$ for the 
U(3)-flavor nonet fields $Y_f$ whose VEVs give effective 
Yukawa coupling constants 
$(Y_f^{eff})_{ij}=\langle (Y_f)_{ij}\rangle/\Lambda$ 
 ($\Lambda$ denotes an energy scale of the effective theory) 
and thereby we obtain relations among masses and mixings from
SUSY vacuum conditions. 
Although, in Ref.\cite{nonetVEV}, 
it has been tried to understand a charged lepton mass relation 
\cite{Koidemass} on the basis of such the approach, 
the purpose of the present paper is not 
to understand such mass spectra, so that 
when we evaluate matrices $\langle Y_u\rangle$ and 
$\langle Y_e\rangle$,  
we will use the observed values of $\langle Y_u\rangle^D 
\propto {\rm diag}(m_u, m_c, m_t)$ and 
$\langle Y_e\rangle^D \propto {\rm diag}(m_e, m_\mu, m_\tau)$ 
($A^D$ denotes a diagonal form of a matrix $A$),
respectively.  
Although our goal is a unified understanding of quark and
lepton mass matrices, in this paper, our investigation will 
focus on the neutrino mass matrix.  
As a result,  we will obtain a neutrino mass matrix
$$
M_\nu = m_0^\nu  \left(\langle Y_e^{-1}\rangle 
\langle Y_u^{1/2} \rangle+\langle Y_u^{1/2}\rangle 
\langle Y_e^{-1}\rangle +
\xi_0 {\bf 1}\right) ,
\eqno(1.1) 
$$
where the charged lepton and up-quark mass matrices 
$M_e$ and $M_u$ are given by $M_e = y_e
\langle Y_e\rangle /\Lambda$ and 
$M_u = y_u\langle Y_u\rangle/\Lambda$.
(For $Y_u^{1/2}$, see Sec.3.)

In order to estimate a neutrino mixing matrix, we must
know an explicit form of (1.1) in a flavor basis in which 
a charged lepton mass matrix $M_e$ is diagonal (we refer 
it as ``$e$-basis").
Especially, we must know an explicit form 
$\langle Y_u^{1/2}\rangle$ 
in Eq.(1.1), although we know the form of $\langle Y_u\rangle$
in the ``$d$-basis" in which a down-quark mass matrix $M_d$
is diagonal, i.e. 
$\langle Y_u\rangle_d$ is given by
$\langle Y_u\rangle_d = V^\dagger\langle Y_u\rangle_u V$,
where $\langle A\rangle_f$ denotes the matrix form of the VEV
matrix $A$ on the $f$-basis, and $V$ is the Cabibbo-Kobayashi-Maskawa 
(CKM) mixing matrix. 
In the present paper, we do not assume a grand unification scenario,
so that the $e$-basis cannot theoretically be related to the $d$-basis.
Nevertheless, we will assume that the form  
$\langle Y_u^{1/2}\rangle_e$ 
is given by a relation
$$
\langle Y_u^{1/2}\rangle_e = V^\dagger(\delta_{ue}) 
\langle Y_u^{1/2}\rangle_u V(\delta_{ue}) ,
\eqno(1.2)
$$
on analogy of $\langle Y_u^{1/2}\rangle_d$, 
where $V(\delta_{ue})$ denotes a mixing matrix in which
the $CP$-violating phase $\delta$ in the CKM matrix $V(\delta)$
is replaced with a free parameter $\delta_{ue}$.
Then, we will find that the numerical results for the neutrino
mass matrix (1.1) with the observed up-quark and charged lepton
masses and CKM matrix parameter (except for $\delta$) can give
a nearly tribimaximal mixing when we take $\delta_{ue} \simeq \pi$.
Of course, there is no theoretical ground in the assumption (1.2),
and it is pure phenomenological one.
Therefore, the neutrino mass matrix which gives a tribimaximal
mixing is also completely empirical one.
Nevertheless, we think that this will provide a promising clue 
to a unification mass matrix model of the quarks and leptons.

In the next section, for convenience of the present investigation, 
we will define an operators $U_{ff'}$ which transforms a matrix 
from a $f$-basis to another $f'$-basis.
In Sec.3, we will assume a form of $W_\nu$ which is composed of
cross terms not only between $Y_\nu$ and $Y_e$, but also
between $Y_e$ and $Y_u^{1/2}$, and thereby, we will discuss 
a neutrino mass matrix of a new type from the phenomenological 
point of view. 
The neutrino mass matrix is described in terms of the up-quark
masses.
Numerical study will be given in Sec.4.
We will find that the  mass matrix (1.1) with the phenomenological
assumption (1.2) can accommodate to a nearly tribimaximal 
mixing \cite{tribi} without
assuming a discrete symmetry, but with assuming an empirical
relation between the $e$- and $d$-bases.
Finally, Sec.5 will be devoted to concluding remarks.

\vspace{3mm}
\noindent{\large\bf 2 \ Flavor-basis transformation}

Note that the matrix form 
$M_e = (y_e/\Lambda)\langle Y_e\rangle$ in the $e$-basis
is diagonal from the definition of the $e$-basis, i.e.
$M_e = (y_e/\Lambda)\langle Y_e\rangle_e = {\rm diag}(m_e, 
m_\mu, m_\tau)$, while the form in another basis is, 
in general, not diagonal. 
Let us begin the present investigation by defining useful
notations on flavor bases.
As we have already used, we define a name of a flavor basis 
as follows:
when a VEV matrix $\langle Y_f\rangle$ takes a diagonal form in a basis,
we refer this basis as ``$f$-basis", and we denote a form 
of a matrix $\langle A \rangle$ on the $f$-basis as 
$\langle A \rangle_f$.
And, we also define a flavor-basis transformation
operator $U_{ab}$ ($a,b=u,d,\nu,e$) by
$$
\langle A \rangle_b =U^\dagger_{ab}\langle A \rangle_{a} U_{ab} ,
\eqno(2.1)
$$
for an arbitrary Hermitian matrix\footnote{
If $Y_f$ are not Hermitian, the definition (2.1) is replaced
with $\langle A \rangle_b =U^\dagger_{Lab}\langle A \rangle_{a} U_{Rab}$.
The assumption that $Y_f$ are Hermitian is not essential
in the present formulation, and the assumption is only
one for convenience.
} $\langle A \rangle$. 
The matrix $U_{ab}$ satisfy the relations $U_{ab}^\dagger= U_{ba}$
and $U_{ab} U_{bc} U_{ca} ={\bf 1}$.
For example, when we assume that $Y_f$ are Hermitian, we can
express 
$$
\langle Y_u \rangle_d = V^\dagger \langle Y_u \rangle_u V \equiv V^\dagger \langle Y_u \rangle^D V ,
\eqno(2.2)
$$
$$
\langle Y_\nu \rangle_e = 
U_\nu \langle Y_\nu \rangle_\nu U_\nu^\dagger \equiv
U_\nu \langle Y_\nu \rangle^D U_\nu^\dagger ,
\eqno(2.3)
$$
where $\langle Y_f \rangle^D$ denote the diagonalized form of 
$\langle Y_f \rangle$,
$V$ is the CKM mixing matrix,
and $U_\nu$ is a neutrino mixing matrix on the $e$-basis
(note that since $Y_\nu$ corresponds to a Dirac mass matrix,
and not to Majorana mass matrix in the seesaw model, the mixing
matrix $U_\nu$ does not always express the observed neutrino 
mixing matrix). 
Therefore, from the definition (2.1), we can regard 
$U_{ud}$ and $U_{e\nu}$ as 
$U_{ud}=V$ and $U_{e\nu}= U_\nu$, respectively.
If the $d$-basis is identical with the $e$-basis, the 
operator $U_{ed}$ will be $U_{ed}={\bf 1}$, so that
$U_{ue}=V$.
In the present paper, we will be interested in whether 
$U_{ed}$ is ${\bf 1}$ or not.
We illustrate our concern in Fig.1.
As we see in Fig.1, if we can determine $U_{ue}$ (or $U_{ed}$)
in addition to the observed $V$ and $U_\nu$,
whole relations among $e$-, $\nu$-, $u$- and $d$-bases can
completely be fixed.
In the present paper, we will search for a possible form
of $U_{ue}$ from the phenomenological point of view.

\begin{figure}[ht]
\begin{center}
\begin{picture}(200,160)(0,0)
\thicklines
\put(0,25){$\langle Y_f \rangle_d$}
\put(140,30){\vector(-1,0){60}}
\put(80,30){\line(-1,0){50}}
\put(145,25){$\langle Y_f \rangle_e$}
\put(0,105){$\langle Y_f \rangle_u$}
\put(140,110){\vector(-1,0){60}}
\put(80,110){\line(-1,0){50}}
\put(145,105){$\langle Y_f \rangle_\nu$}
\put(10,40){\vector(0,1){30}}
\put(10,70){\line(0,1){30}}
\put(150,40){\vector(0,1){30}}
\put(150,70){\line(0,1){30}}
\put(30,90){\vector(2,-1){50}}
\put(80,65){\line(2,-1){40}}
\put(35,5){$U_{ed}=\ ?$}
\put(35,125){$U_{\nu u}=\ ?$}
\put(-60,70){$U_{du}=V^\dagger$}
\put(160,70){$U_{e\nu}= U_\nu$}
\put(70,75){$U_{ue}=\ ?$}
\end{picture}
\end{center}

\begin{quotation}
\caption{Illustration of our concern for flavor-basis 
transformation operators $U_{ab}$, which
is defined by $\langle Y_f \rangle_{b} = U^\dagger_{ab} 
\langle Y_f \rangle_{a} U_{ab}$.
We will search for a possible form of $U_{ue}$ from the
phenomenological point of view.
}
\end{quotation}
\label{fig1}
\end{figure}
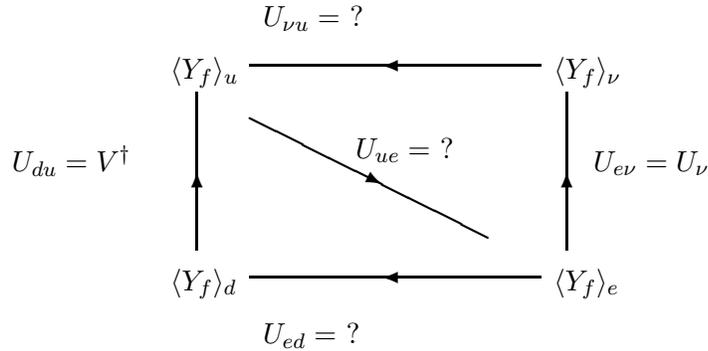

Thus, we try to build a mass matrix model not by investigating
explicit structures of the Yukawa coupling constants
(for example, with assuming some discrete symmetries), but 
by investigating a superpotential for U(3)-flavor nonet fields. 
Such a prescription seems to provide a new approach to quark
and lepton masses and mixings.
The purpose of the present paper is to investigate a possible
structure of $W_\nu$ from the phenomenological point of view.

\vspace{3mm}
\noindent{\large\bf 3 \ SUSY vacuum approach}

In the present approach, the Yukawa coupling constants
are understood as ``effective" coupling constants
$\langle Y_f\rangle/\Lambda$ from the following interactions:
$$
W_{Y}= \sum_{i,j} \frac{y_u}{\Lambda}(Y_u)^i_{j} {Q}_{i} H_u U^{j} 
+\sum_{i,j}\frac{y_d}{\Lambda} (Y_d)^i_{j} {Q}_{i} H_d D^{j} 
$$
$$
+\sum_{i,j} \frac{y_\nu}{\Lambda}(Y_\nu)^i_{j} {L}_{i} H_u N^{j} 
+\sum_{i,j}\frac{y_e}{\Lambda} (Y_e)^i_j {L}_{i} H_d E^{j} 
+h.c. + y_R \sum_{i,j} N^i (M_R)_{ij} N^j ,
\eqno(3.1)
$$ 
where $Y_f$  ($f=u,d,\nu,e$) are not coupling constants, but U(3)-flavor
nonet fields \cite{nonetVEV}, 
and $Q$ and $L$ are quark and lepton 
SU(2)$_L$ doublet fields, respectively, and $U$, $D$, $N$, and $E$ are 
SU(2)$_L$ singlet matter fields.
The mass parameter $\Lambda$ denotes an energy scale of the effective theory.
Therefore, the quark and lepton mass matrices $M_f$ are given by
$$
M_f = \frac{y_f}{\Lambda} \langle Y_f \rangle \langle H^0_{u/d}\rangle ,
\eqno(3.2)
$$
where  $\langle Y_f \rangle/\Lambda \sim 1$.
We consider that the VEVs of $Y_f$ are completely determined 
by another U(3)$_F$ nonet fields, $\Phi_e$ and $\Phi_u$, as we show latter.
For simplicity, we investigate a case that all VEV matrices $Y_f$
($f=u,d,\nu,e$) are Hermitian.

In the present model, the superpotential $W$ is given by
$$
W=W_Y +W_e +W_u +W_\nu +W_d ,
\eqno(3.3)
$$
where $W_f$ ($f=u,d,\nu,e$) play a role in fixing the VEV structure
$\langle Y_f\rangle$.
Since we can easily show $\langle Q\rangle =\langle L\rangle 
=\langle U\rangle =\langle D\rangle =\langle N\rangle 
=\langle E\rangle =0$, hereafter, we will
drop the term $W_Y$ from (3.3) when we investigate the VEV 
structures of $Y_f$. 
Since we focus on the neutrino mass matrix, we will discuss
only $W_u$ and $W_\nu$.  

We assume the following structure of $W_u$ by introducing 
U(3)-nonet fields $Y_u$ and $\Phi_u$:
$$
W_u = \lambda_u {\rm Tr}[\Phi_u\Phi_u\Phi_u] +m_{u}{\rm Tr}[\Phi_u\Phi_u]
+\mu_u^2 {\rm Tr}[\Phi_u]
+\lambda_{Yu} {\rm Tr}[\Phi_u\Phi_u Y_u] +m_{Y} {\rm Tr}[Y_u Y_u] .
\eqno(3.4)
$$
(Although the similar form was assumed for $W_e$ in 
Ref.\cite{nonetVEV},
we will not mention the explicit form of $W_e$ in the present model, 
 because the purpose of the present paper is not to derive 
the charged lepton mass relation as in Ref.\cite{nonetVEV}.)
From the SUSY vacuum conditions, we obtain
$$
\frac{\partial W}{\partial \Phi_u}=3 \lambda_u \Phi_u\Phi_u + 2 m_{u}\Phi_u +\mu_u^2 {\bf 1}
+\lambda_{Yu} (\Phi_u Y_u+Y_u\Phi_u) = 0 ,
\eqno(3.5)
$$
$$
\frac{\partial W}{\partial Y_u} =
\lambda_{Yu} \Phi_u\Phi_u+ 2 m_{Yu} Y_u = 0 .
\eqno(3.6)
$$
(For the moment, we take $W=W_u$.)
Therefore, we can obtain a bilinear mass relation 
$$
 Y_u = -\frac{\lambda_{Yu}}{2 m_{Yu}} \Phi_u \Phi_u ,
\eqno(3.7)
$$
from Eq.(3.6).
The operator $Y_u^{1/2}$ in Eq.(1.1) means the present operator
$\Phi_u$.  Hereafter, we use $\Phi_u$ instead of $Y_u^{1/2}$.
On the other hand, 
by substituting Eq.(3.7) into Eq.(3.5), we obtain
$$
c_3 \Phi_u\Phi_u\Phi_u + c_2 \Phi_u\Phi_u + 
c_1 \Phi_u + c_0 {\bf 1} =0 ,
\eqno(3.8)
$$
where $c_3={\lambda_{Yu}^2}/{m_{Yu}}$, $c_2 = -3\lambda_u$, 
$c_1 = - 2m_{u}$ and $c_0 = -\mu_u^2$.
Thus, if we give values of the coefficients $c_n$  ($n=3,2,1,0$),
we can completely determine three eigenvalues of $\langle\Phi_u\rangle$,
so that we can also completely determine three eigenvalues of 
$\langle Y_u\rangle$. 
We assume that the superpotential (3.3) does not include 
any explicit flavor symmetry breaking parameter.
The most distinctive feature of the present model is 
that the U(3) flavor symmetry is spontaneously and completely 
broken by the non-zero and non-degenerate VEVs of 
$\langle\Phi_e\rangle$, 
without passing any subgroup of U(3)$_F$. 
(For example, differently from the present model,
a U(3)$_F$-nonet scalar $\Phi$ in Ref.\cite{YK-JHEP07} is
broken, not directly, but via a discrete symmetry S$_4$.)

Next, we investigate a possible form of $W_\nu$ which
leads to phenomenologically successful neutrino mass matrix.
For the moment, we neglect the term $W_d$ from Eq.(3.3), i.e.
we regard $W$ as $W=W_e +W_u +W_\nu$.
We suppose that the matrix $Y_\nu$ will be related to
the up-type VEV matrix $\Phi_u$ by considering a correspondence
$Y_e \leftrightarrow Y_d$ and $Y_\nu \leftrightarrow Y_u$.
However, if $Y_\nu$ is described in terms of $\Phi_u$ and $Y_u$
only, the matrix $Y_\nu$ is also diagonalized
on the $u$-basis as well as $Y_u$ and $\Phi_u$.
Since the observed neutrino mixing matrix is peculiarly
different from the observed CKM matrix structure, we
must consider that the $\nu$-basis is different from
the $u$-basis.
Therefore, we consider that $W_\nu$ is a function not only of
$Y_u$ and $\Phi_u$, but also of $Y_e$ (and/or $Y_d$).   
By way of trial, we assume the following form of $W_\nu$:
$$
W_\nu = \frac{y_\nu }{\Lambda} {\rm Tr}[Y_e Y_\nu Y_e \Phi_0]
+\lambda_{\nu 1} {\rm Tr}[(Y_e \Phi_u +\Phi_u Y_e)\Phi_0] 
+ \lambda_{\nu 2} {\rm Tr}[Y_e Y_e \Phi_0] ,
\eqno(3.9)
$$ 
where the new nonet field $\Phi_0$ has been introduced in order 
that SUSY vacuum conditions for $W_\nu$ do not change 
relations (3.7) and so on, which are derived from SUSY vacuum 
conditions for $W_e$ and $W_u$.
(In the form (3.9), it is not a general form of possible terms
which include $Y_\nu$. Our concern is what specific form of
$W$ can lead to a successful phenomenology, and not what 
principle can lead to such a specific form of $W$.)
From the SUSY vacuum condition 
$$
\frac{\partial W}{\partial Y_\nu} = 0 = \frac{y_\nu }{M} 
Y_e \Phi_0 Y_e ,
\eqno(3.10)
$$
we obtain
$$
\langle\Phi_0\rangle =0 ,
\eqno(3.11)
$$
for $Y_e\neq 0$, so that we obtain
$$
\frac{\partial W}{\partial Y_e} = 0 = 
\lambda _{Ye} \Phi_e \Phi_e + 2 m_{Ye} Y_e 
+ \frac{y_\nu }{\Lambda} (Y_\nu Y_e \Phi_0 +\Phi_0 Y_e Y_\nu) 
$$
$$
+\lambda_{\nu 1} (\Phi_u \Phi_0+ \Phi_0 \Phi_u) 
+\lambda_{\nu 2} (Y_e \Phi_0+ \Phi_0 Y_e) 
=\lambda _{Ye} \Phi_e \Phi_e + 2 m_{Ye} Y_e .
\eqno(3.12)
$$
On the other hand, from $\partial W/\partial \Phi_0=0$,
we obtain
$$
\frac{\partial W}{\partial \Phi_0} = 0 = 
\frac{y_\nu }{\Lambda} Y_e Y_\nu Y_e  + \lambda_{\nu 1} (Y_e \Phi_u
+\Phi_u Y_e) + \lambda_{\nu 2} Y_e Y_e ,
\eqno(3.13)
$$
i.e.
$$
\frac{y_\nu }{\Lambda} Y_\nu =- \lambda_{\nu 1} (Y_e^{-1} \Phi_u +\Phi_u Y_e^{-1} )
-\lambda_{\nu 2} {\bf 1} .
\eqno(3.14)
$$
The relation (3.14) means
$$
(M_\nu^{Dirac})_{ij} = m_0^\nu \left[ 
\left( \frac{1}{m_{ei}} +  \frac{1}{m_{ej}}
\right) (\langle\Phi_u\rangle_e)_{ij}  + \xi_0 \delta_{ij} \right] ,
\eqno(3.15)
$$
where $m_{ei}$ are the charged lepton masses.
From the definition (2.1) of the flavor-basis transformation, 
the form of $\langle\Phi_u\rangle_e$ is expressed by
$$
\langle\Phi_u\rangle_e = U_{ue}^\dagger \langle\Phi_u\rangle_u U_{ue}
=v_u U_{ue}^\dagger Z_u U_{ue} ,
\eqno(3.16)
$$
where 
$\langle \Phi_u\rangle = v_u Z_u \equiv 
v_u{\rm diag}(z_1^u, z_2^u, z_3^u)$ and
$$
z_i^u = \frac{\sqrt{m_{ui}}}{\sqrt{m_{u1}+m_{u2}+m_{u3}}} ,
\eqno(3.17)
$$
($m_{ui}$ are up-quark masses) from Eq.(3.7). 

The mass matrix (3.15) has a very peculiar form because
the matrix includes up-quark masses ($(\langle\Phi_u\rangle^D)_{ii}
\propto \sqrt{m_{ui}}$).
If we can know a form of $U_{ue}$, we can obtain an 
explicit form of the Dirac neutrino mass matrix (3.15) 
except for the common shift term ($\xi_0$-term), so that 
we can calculate the mixing matrix $U_\nu$
independently of the value of the parameter $\xi_0$.
However, at this stage, the form (3.15) does not have any 
theoretical basis.
Moreover, we have no principle to decide the form of $U_{ue}$.
In the next section, we will investigate the mass matrix 
(3.15) from the phenomenological point of view, and we 
will demonstrate that the mass matrix (3.15) can give a nearly
tribimaximal mixing when we assume a simple specific form
of $U_{ue}$.


\vspace{3mm}
\noindent{\large\bf 4 \ Phenomenological investigation of 
the neutrino mass matrix}

In the present section, we assume a form of $U_{ue}$, and thereby, we 
investigate the mass matrix (3.15) from the phenomenological 
point of view.

\vspace{2mm}
\noindent{\bf 4.1 \ Numerical study of the Dirac neutrino mass matrix}

First, we investigate a case that the observed neutrinos are
Dirac type and the mass matrix is given by (3.15).
The simplest assumption for a form of $U_{ue}$ is to consider 
that the $d$-basis is identical with the $e$-basis, so that 
we can regard $U_{ue}$ as $U_{ue} =V$ because $U_{ud}= V$.
Then, we can evaluate the form $M_\nu^{Dirac}$ except for the
common shift term $\xi_0$, so that we can obtain the neutrino
mixing angles $\sin^2 2\theta_{23}$ and $\tan^2\theta_{12}$.
The numerical results are shown in Table 1 for the following
input values: the up-quark masses \cite{FK-qmass} at $\mu=M_Z$,  
$m_{u1}=0.00233$ GeV, $m_{u2}=0.677$ GeV, $m_{u3}=181$ GeV,
 and the CKM parameters \cite{PDG06},
$|V_{us}|=0.2257$, $|V_{cb}|=0.0416$, $|V_{ub}|=0.00431$.
(Here, we have used the quark mass values at $\mu=M_Z$ because
we have used the CKM parameter values at $\mu=M_Z$.  For the
energy scale dependency of the mass ratios and CKM parameters,
for example, see Ref.\cite{evol}.)
The standard phase convention \cite{PDG06} has been adopted as a phase
convention of $V$.  
The present experimental data \cite{PDG06} show $\delta \simeq \pi/3$.   
However, as seen in Table 1, the predicted values of 
$\sin^2 2\theta_{23}$ and $\tan^2\theta_{12}$ at  $\delta \simeq \pi/3$
are in poor agreement with the observed values.
Of course, the mixing matrix $U_\nu$ defined (2.3) is one for
the Dirac neutrino matrix, it is not observed one if the observed
neutrinos are Majorana type.
However, when we take a seesaw neutrino mass matrix model,
the predictions of $\sin^2 2\theta_{23}$ and 
$\tan^2\theta_{12}$ at  $\delta \simeq \pi/3$ become 
all the more worse (even adjusting the parameter $\xi_0$)
as we see later (in Table 2).
Therefore, we cannot regard that the $d$-basis is identical with
the $e$-basis, i.e. in other words, the matrix $Y_d$ cannot
simultaneously be diagonalized together with $Y_e$. 
We cannot regard $U_{ue}$ as $U_{ue}=V$.

\begin{table}

\begin{quotation}
\caption{
$\delta_{ue}$ dependence of the neutrino Dirac mass matrix
$Y_\nu$.  
The numerical values of $(M_\nu^{Dirac})_{ij}$ are given in unit of 
$m_0^\nu v_u/m_0^e$
in Eq.(3.15) for the case of $\xi_0=0$.
The values of $\sin^2 2\theta_{23}$ and $\tan^2 \theta_{12}$ are 
estimated by 
$\sin^2 2\theta_{23} = 4 |(U_\nu)_{23}|^2 |(U_\nu)_{33}|^2/(1-|U_{13}|^2)$ and
$\tan^2 \theta_{12}=|(U_\nu)_{12}/(U_\nu)_{11}|^2$, 
respectively.
}
\end{quotation}

\begin{center}
\begin{tabular}{|c|l|l|l|l|c|c|c|}\hline
$\delta_{ue}$ & $(M_\nu)_{22}$ &  $(M_\nu)_{33}$ &
$(M_\nu)_{12}$ & $(M_\nu)_{13}$ & $\sin^2 2\theta_{23}$ &
$\tan^2 \theta_{12}$ & $|U_{13}|$ \\ \hline
0    & $0.1579$  & $0.1568$ & $-3.526$ & $1.264$ & $0.3831$ & $0.4170$
& $0.0113$ \\ 
$60^\circ$ & $0.1579$ & $0.1568$ & $-3.547 e^{i0.65^\circ}$ &
$2.083 e^{i 28.3^\circ}$ & $0.7545$ & $0.4477$ & $0.0085$ \\
$90^\circ$ & $0.1577$ & $0.1568$ & $-3.568 e^{i 0.074^\circ}$ &
$2.660 e^{i 25.4^\circ}$ & $0.9159$ & $0.4730$ & $0.0061$ \\
$120^\circ$ & $0.1577$ & $0.1568$ & $-3.589 e^{-i 0.64^\circ}$ &
$3.134 e^{-i 18.4^\circ}$ & $0.9813$ & $0.4943$ & $0.0039$ \\
$180^\circ$ & $0.1576$  & $0.1568$ & $-3.609$ & $3.544$ & $0.9997$ 
& $0.5125$ & $0.0001$ \\ \hline
\end{tabular}\end{center} 
\end{table}

We still expect that $U_{ue} \simeq U_{ud}$, i.e. 
$U_{ed}\simeq {\bf 1}$.
Therefore, next, we investigate a possibility of
$U_{ue}=V(\delta_{ue})$, where $V(\delta)$ is the
standard expression of the CKM mixing matrix $V$ with the $CP$
violating phase $\delta$.
The observed data \cite{PDG06} on the CKM matrix parameters show
$\delta \simeq \pi/3$.
For simplicity, hereafter, we will regard the CKM matrix $V$ as $V(\pi/3)$,
and for $U_{ue}$, we will denote $U_{ue}=V(\delta_{ue})$, where
we regard $\delta_{ue}$ as a free parameter.
Then, we can show 
$$
U_{ed}= U_{ue}^\dagger U_{ud}=V^\dagger(\delta_{ue}) V(\frac{\pi}{3})
={\bf 1} - \left(
\begin{array}{ccc}
\varepsilon_{11} & \varepsilon_{12} & \varepsilon_{13} \\
\varepsilon_{12}  & \varepsilon_{22} & \varepsilon_{23} \\
-\varepsilon_{13}^* & -\varepsilon_{23}^* & \varepsilon_{33}
\end{array} \right)
  = {\bf 1} +{\cal O}(|V_{ub}|) ,
\eqno(4.1)
$$
where $\varepsilon_{11} = (1-e^{-i(\delta_{ue}-\pi/3)})
c_{12}^2 s_{13}^2$,  
$\varepsilon_{22}  = (1-e^{-i(\delta_{ue}-\pi/3)}) 
s_{12}^2 s_{13}^2 $,
$\varepsilon_{33}  = (1-e^{i(\delta_{ue}-\pi/3)}) s_{13}^2 $,
$\varepsilon_{12}  = (1-e^{-i(\delta_{ue}-\pi/3)}) 
s_{12}c_{12} s_{13}^2$, 
$\varepsilon_{13}  = (e^{-i\delta_{ue}} -e^{-i\pi/3})
c_{12} s_{13} c_{13}$,  and 
$\varepsilon_{23}  = (e^{-i\delta_{ue}} -e^{-i\pi/3})
 s_{12} s_{13} c_{13}$ ($s_{ij}=\sin\theta_{ij}$ and 
$c_{ij}\cos\theta_{ij}$ are rotation parameters in the 
standard phase convention of the CKM matrix \cite{PDG06}). 

As shown in Table 1, the cases $U_{ue}=V(\delta_{ue})$ with
$(2/3)\pi \leq |\delta_{ue}| \leq \pi$ can give a reasonable set
of $(\sin^2 2\theta_{23}, \tan^2 \theta_{12})$
for the observed values $\tan^2\theta_{12}=0.47^{+0.06}_{-0.05}$
\cite{KamLAND} and $\sin^2 2\theta_{23}=1.00_{-0.13}$
\cite{MINOS}.
Especially, we are interested in the cases, 
(i) $\delta_{ue}=\pi +\pi/3$ and (ii) $\delta_{ue}=\pi$.
The case (i) gives $\sin^2 2\theta_{23}=0.981$ and
$\tan^2 \theta_{12}=0.494$, and it is likely that the form 
$U_{ue} =V(\pi +\delta_{CKM})$ can be understood in a future 
theoretical model.
On the other hand, the case (ii) is also interesting, because
the case can give a mixing highly close to the so-called
tribimaximal mixing \cite{tribi} (i.e. the case gives
$\sin^2 2\theta_{23}=1.000$ and $\tan^2 \theta_{12}=0.513$),
and the mixing matrix $U_{ue}$ is an orthogonal matrix
(it does not include phase parameters),  so that $U_\nu$
is also an orthogonal one.

More precisely speaking, the tribimaximal mixing takes place
only when $(Y_\nu)_{22}=(Y_\nu)_{33}$ and
$(Y_\nu)_{12} =\pm (Y_\nu)_{13}$.
In the present model, 
the ratio $(Y_\nu)_{22}/(Y_\nu)_{33} \simeq 1$ is 
satisfied for any value of $\delta_{ue}$ in 
$U_{ue}=V(\delta_{ue})$.
This is warranted by the observed fact
$$
\frac{m_\mu}{m_\tau} \simeq \sqrt{ \frac{m_c}{m_t}} .
\eqno(4.2)
$$
On the other hand, the ratio $(Y_\nu)_{12}/(Y_\nu)_{13}$ is 
highly sensitive to the value of $\delta_{ue}$, because
$$
\frac{(Y_\nu)_{12}}{(Y_\nu)_{13}}
\simeq \frac{V_{21}^* V_{22} \sqrt{m_c} +\cdots}{V_{31}^* V_{33}
\sqrt{m_t} +\cdots}
\simeq  -\frac{|V_{us}|}{V_{31}^*} \sqrt{\frac{m_c}{m_t}},
\eqno(4.3)
$$
$$
V^*_{31}(\delta_{ue}) \simeq |V_{us}| |V_{cb}| -|V_{ub}| e^{-i \delta_{ue}}.
\eqno(4.4)
$$
The relation $(Y_\nu)_{12}/(Y_\nu)_{13} \simeq -1$
with $\delta_{ue}=\pi$ is warranted by the fact that the relation
$$
\sqrt{ \frac{m_c}{m_t} } \simeq |V_{cb}| + 
\frac{|V_{ub}|}{|V_{us}|} ,
\eqno(4.5)
$$
is well satisfied with the observed values,
$\sqrt{ {m_c}/{m_t} }=0.061$  \cite{FK-qmass} at $\mu=M_Z$, 
$|V_{cb}|=0.0416$ and
$|V_{ub}|/|V_{us}|=0.0191$ \cite{PDG06}.

\vspace{2mm}

\noindent{\bf 4.2 \ Numerical study of the seesaw neutrino mass matrix}

Next, we investigate a case that the observed neutrinos are 
Majorana neutrinos which are generated by a seesaw mechanism:
$$
M_\nu =\left( \frac{y_\nu}{\Lambda} v_{Hu}  \right)^2
Y_\nu M_R^{-1} Y_\nu^T.
\eqno(4.6)
$$
In this case, the mixing matrix $U_\nu$ is not always identical with 
a mixing matrix $U_{e\nu}$ which is defined as
$$
U^\dagger_{e\nu} \langle Y_\nu\rangle_e U_{e\nu} 
= \langle Y_\nu\rangle_\nu \equiv \langle Y_\nu\rangle^D.
\eqno(4.7)
$$
In order to diagonalize the matrix (4.6), we must know a
form of $\langle M_R\rangle_e$.
For simplicity, we assume that the form of $M_R$ is 
independent of the flavor basis, i.e. $M_R\propto {\bf 1}$.
Then, the mixing matrix $U_\nu$ is obtained by diagonalizing 
the matrix $\langle Y_\nu\rangle_e \langle Y_\nu\rangle_e^T$.
When we denote $\langle Y_\nu\rangle_e$ as 
$\langle Y_\nu\rangle_e = Y_0 +\xi {\bf 1}$, where
$U_{e\nu}^\dagger Y_0 U_{e\nu}=D_0$ ($D_0$ is diagonal),
we can show 
$U_{e\nu}^\dagger \langle Y_\nu\rangle_e \langle Y_\nu\rangle_e^T
U_{e\nu}^* = (D_0 +\xi {\bf 1}) U_{e\nu}^\dagger U_{e\nu}^*
(D_0 +\xi {\bf 1})$.
Since the transformation matrix $U_{e\nu}$ is orthogonal in the
cases with $\delta_{ue}=0$ and $\delta_{ue}=\pi$, the matrix 
$U_{e\nu}^\dagger U_{e\nu}^*$ becomes a unit matrix ${\bf 1}$,
so that the lepton mixing matrix $U_\nu$ is given by
$U_\nu = U_{e\nu}$ as well as in the case of Dirac neutrinos.
However, for the cases with $\delta_{ue} \neq 0$ and 
$\delta_{ue} \neq \pi$,
the case of Majorana neutrinos cannot give the same results with
the case of Dirac neutrinos. 
The numerical results are given in Table 2.
Although, in the case of Dirac neutrinos, the case with
$\delta_{ue}=\delta_{CKM} +\pi$ ($\simeq -120^\circ$) has been
acceptable, in the present case, such a case with 
$\delta_{ue} \neq \pi$ is ruled out, because such case can give
reasonable values for neither $\tan^2\theta_{12}$ nor 
$R \equiv \Delta m^2_{21}/\Delta m^2_{32}$ as seen in Table 2.
(Also note that the value of $\tan^2\theta_{12}$ is
sensitive to the value of $\xi_0$.)

\begin{table}

\begin{quotation}
\caption{
Dependence on $\delta_{ue}$ and $\xi_0$ for the predicted values
$R \equiv \Delta m^2_{21}/\Delta m^2_{32}$, $\sin^2 2\theta_{23}$, 
$\tan^2 \theta_{12}$ and $|U_{13}|$ in the seesaw mass matrix (4.6). 
The values of $R$ in the case with $\delta_{ue}=0$ and
$\delta_{ue}=\pi$ have already been adjusted by
fitting $\xi_0$ to the value $|R| \simeq 0.028$ and 
$\Delta m^2_{21} >0$.
For the cases $\delta_{ue}=\pi/3$ and $2\pi/3$, 
we cannot obtain such a small value  
as $|R| \simeq 0.028$, so that we denote only the cases with lower
limits of $R$ in Table.
}
\end{quotation}

\begin{center}
\begin{tabular}{|r|c|r|c|c|c|}\hline
$\delta_{ue}$ & $\xi_0/(v_u/m_0^e)$ & 
$R $ & $\sin^2 2\theta_{23}$
& $\tan^2 \theta_{12}$ & $|U_{13}|$ \\ \hline
$0^\circ$  & $-25.57$ & $-0.0282$ & $0.3831$ &  $0.4170$  
& $0.0113$ \\
$60^\circ$ & $-25.85$ & $ -0.1170$ & $0.7035$ & $0.0757$ 
& $0.1349$ \\
$60^\circ$ & $-25.99$ & $ -0.1156$ & $0.7032$ & $0.0381$ 
& $0.1355 $ \\
$60^\circ$ & $-26.13$ & $ -0.1168$ & $0.7029$ & $0.0142$ 
& $0.1360 $ \\
$120^\circ$ & $-25.99$ & $ -0.0705$  & $0.9643$ & $0.0889$ 
& $0.0912$ \\
$120^\circ$ & $-26.13$ & $ -0.0685$  & $0.9641$ & $0.0324$ 
& $0.0912$ \\
$120^\circ$ & $-26.27$ & $ -0.07000$  & $0.9640$ & $0.0054$ 
& $0.0912$ \\
$180^\circ$  & $-26.28$ & $-0.0273$ &  $0.9997$ & $0.5125$
& $0.0001$ \\
\hline
\end{tabular}
\end{center}
\end{table}

Thus, if we consider that the observed neutrinos are  
Majorana types, only the case $U_{ue}=V(\pi)$ can give successful predictions:
$$
U_\nu = \left(
\begin{array}{rrr}
0.8131 & -0.5821 & -0.0001 \\
-0.4153 & -0.5803 & 0.7006 \\
0.4079 & 0.5696 & 0.7136 
\end{array} \right) ,
\eqno(4.8)
$$
independently of the value of $\xi_0$.
The result (4.8) is very close to the tribimaximal 
mixing
$$
U_{TB} = \left( 
\begin{array}{ccc}
\frac{2}{\sqrt6} &  -\frac{1}{\sqrt3} & 0 \\
-\frac{1}{\sqrt6} &  -\frac{1}{\sqrt3} & \frac{1}{\sqrt2} \\
\frac{1}{\sqrt6} &  \frac{1}{\sqrt3} & \frac{1}{\sqrt2} 
\end{array} \right) .
\eqno(4.9)
$$
(Here, we have taken a phase convention corresponding to (4.8).)
We simply regard that $U_{e\nu}=U_{TB}$.
The relations among four bases are illustrated in Fig.~2.

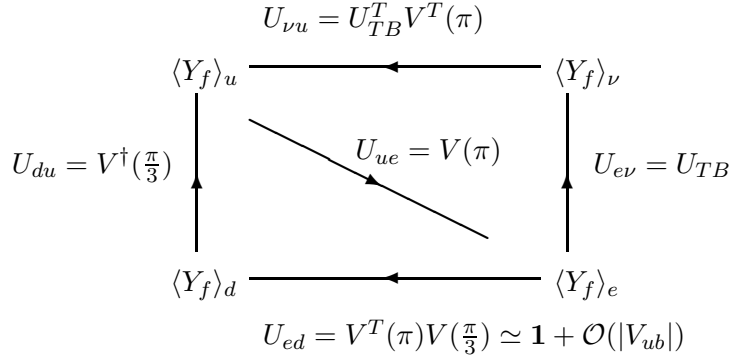
\begin{figure}[ht]
\begin{center}
\begin{picture}(200,160)(0,0)
\thicklines
\put(0,25){$\langle Y_f\rangle_d$}
\put(140,30){\vector(-1,0){60}}
\put(80,30){\line(-1,0){50}}
\put(145,25){$\langle Y_f\rangle_e$}
\put(0,105){$\langle Y_f\rangle_u$}
\put(140,110){\vector(-1,0){60}}
\put(80,110){\line(-1,0){50}}
\put(145,105){$\langle Y_f\rangle_\nu$}
\put(10,40){\vector(0,1){30}}
\put(10,70){\line(0,1){30}}
\put(150,40){\vector(0,1){30}}
\put(150,70){\line(0,1){30}}
\put(30,90){\vector(2,-1){50}}
\put(80,65){\line(2,-1){40}}
\put(35,5){$U_{ed}=V^T(\pi) V(\frac{\pi}{3})\simeq {\bf 1}+{\cal O}(|V_{ub}|)$}
\put(35,125){$U_{\nu u}=U_{TB}^T V^T(\pi)$}
\put(-60,70){$U_{du}=V^\dagger(\frac{\pi}{3}) $}
\put(160,70){$U_{e\nu}= U_{TB}$}
\put(70,75){$U_{ue}= V(\pi)$}
\end{picture}
\end{center}

\begin{quotation}
\caption{Flavor-basis transformation operators $U_{ab}$, which
is defined by $\langle Y_f\rangle_{b} = U^\dagger_{ab} 
\langle Y_f\rangle_{a} U_{ab}$
($a,b,c=u,d,\nu,e$).
}
\end{quotation}

\label{fig1}
\end{figure}

In the present model, the absolute values of the neutrino masses
cannot be predicted because of the free parameter $\xi_0$.
We can merely choose a suitable value of $\xi_0$ from
the observed value
$$
|R|=\frac{\Delta m^2_{21}}{|\Delta m^2_{32}|}
=0.028\pm 0.004 ,
\eqno(4.10)
$$
where we have used the observed values 
$\Delta m^2_{21}=(7.59\pm 0.21) \times 10^{-5}\, {\rm eV}^2$
\cite{KamLAND} and 
$|\Delta m^2_{32}|=(2.74^{+0.44}_{-0.26}) \times 10^{-3}\, {\rm eV}^2$
\cite{MINOS}.
The numerical results are demonstrated in Table 3, where
the values of $\xi_0$ are given in unit of $v_u/m_0^e$,
$\langle(\Phi_u^D)_{ii}\rangle =v_u z_i^u \propto \sqrt{m_{ui}}$
and $m_{ei}= m_0^e (z_i^e)^2$, where $z_i^e$ and $z_i^u$ are
normalized as $z_1^2+z_2^2+z_3^2=1$.
Note that the present model gives an inverted mass hierarchy of 
neutrinos.
The values of $m_{\nu i}$ are estimated by putting as
$m_{\nu 2}= \sqrt{|\Delta m^2_{32}|}=0.0523$ eV.
As seen in Table 3, the numerical results $\sum m_{\nu i} =0.12$
eV (Dirac neutrinos) and $\sum m_{\nu i} =0.11$ eV (Majorana
neutrinos) safely satisfies the cosmological lower bound 
$\sum m_{\nu i} < (0.2-0.4)$ eV (the recent cosmological
neutrino mass bounds are listed, for example, in Ref.\cite{sum_mnu}).
For a case of Majorana neutrinos, we can calculate the effective
neutrino mass $\langle m_{ee}\rangle $ as
$$
|\langle m_{ee} \rangle | =\left|\sum_i U_{ei}^2 m_{\nu i}\right|  
= 0.0164 \ {\rm eV}.
\eqno(4.11)
$$
The value (4.11) will be observed in future neutrinoless double
beta experiments. 

\begin{table}

\begin{quotation}
\caption{
Neutrino masses fitted from the observed values 
$R=\Delta m^2_{21}/\Delta m^2_{32}$ and
$m_{\nu 2}=\sqrt{|\Delta m^2_{32}|}$. 
For seesaw masses, the Majorana mass matrix $M_R$ of 
the right-handed neutrinos is assumed as $M_R \propto {\bf 1}$.
}
\end{quotation}

\begin{center}
\begin{tabular}{|c|c|c|c|c|c|c|}\hline
Type & $\xi_0/(v_u/m_0^e)$ & $R$ & $m_{\nu 1}$ [eV] 
& $m_{\nu 2}$ [eV] & $m_{\nu 3}$ [eV]  & $\sum m_{\nu i}$ [eV] 
\\ \hline
Dirac mass & $-26.48$ & $-0.0276$ & $0.0517$ & $0.0523$ & $0.0181$ 
& $0.1221$ \\
Seesaw mass & $-26.29$ & $-0.0281$ & $0.0516$ & $0.0523$ 
& $0.0062$  & $0.1101$ \\ \hline
\end{tabular}
\end{center}
\end{table}


\vspace{3mm}
\noindent{\large\bf 5 \ Concluding remarks}

In conclusion, based on a U(3)-flavor nonet scalar model, 
we have obtained a neutrino mass matrix (3.15) of a new type,
where the matrix is described in terms of charged lepton  and 
up-quark mass matrices.
However, in order to evaluate the neutrino mixing matrix
from the neutrino mass matrix (3.15), we must know the form
of $\langle Y_u \rangle$ on the $e$-basis (not on the $d$-basis).
Since we do not know it at present, we have assumed the form (1.2)
from the phenomenological point of view.
Then, we have found the neutrino mass matrix (3.15) with the 
phenomenological assumption (1.2) can give a nearly tribimaximal
mixing.
(Therefore, as shown in the present title, the neutrino mass 
matrix is not one which is derived from a model, but it is 
an empirical one.)
Nevertheless, it is worthwhile noticing because the form is
one of a new type which is related to the up-quark masses and
which successfully leads to the nearly tribimaximal mixing
without assuming any discrete symmetry. 
Inversely speaking, this phenomenological success suggests
a possibility that we can understand the CKM matrix and quark
mass spectrum by starting from a discrete symmetry which 
gives the observed tribimaximal mixing for the lepton sectors.   

If we accept the empirical neutrino mass matrix (3.15),  
in order that the neutrino mass matrix $Y_\nu$ gives 
successful results, we cannot regard that the $e$-basis
is identical with the $d$-basis, and we must take
$U_{ue}=V(\delta_{ue})$ with $2 \pi/3 \leq \delta_{ue} \leq \pi$
for Dirac neutrinos and with $\delta_{ue}=V(\pi)$ for the 
seesaw (Majorana) neutrinos, 
although the $e$-basis is still very near to the $d$-basis, i.e.
$U_{ed}={\bf 1}+{\cal O}(|V_{ub}|)$, Eq.(4.1).
The present model gives an inverse hierarchy of the neutrino
masses as seen in Table 3.
The reason why $U_{ue}$ takes the form 
$V(\delta_{ue})$ is, at present, an open question, and
it is only a phenomenological conclusion.

In this paper, we have not investigate a possible form
of $W_d$ which will give relations of the field $Y_d$ to
other fields\footnote{
For example, in Ref.\cite{nonetVEV}, a model for $W_d$
has been proposed.  However, in the model, since the $d$-basis 
is identical with the $e$-basis, we cannot apply the model
to the present model straightforwardly. 
}.  
Since we have given $U_{ue}$, the relative relations 
among four flavor-bases are fixed each other.
Therefore, if we give a form of $W_d$, we can give 
not only an explanation of the down-quark masses, but also 
``predictions" for other masses and mixings.
However, in order to give an explicit form of $W_d$,
we must put further assumptions, so that we have not
discussed the explicit form of $W_d$ because the purpose
of the present paper is to report an empirical neutrino
mass matrix of a new type.
A possible model for $Y_d$ will be given elsewhere.

By the way, we have not discussed a possibility that
the present model is extended to a grand unification (GUT) 
scenario. 
In the present model, since all $Y_f$ are assumed as 
U(3)-flavor nonets, the model cannot be extended to
GUT scenario, because $Y_u$, for example, will be a 6-plet
of U(3)$_F$ in a GUT model, because ${\bf 3}\times {\bf 3}=
{\bf 6}_S +{\bf 3}^*_A$ (not ${\bf 3}\times {\bf 3}^*=
{\bf 1} +{\bf 8}$).
When $Y_u$ is a 6-plet of U(3)$_F$, it is hard to lead such 
a bilinear relation as Eq.(3.7).
If we want a formulation similar to the present prescription,
we may consider, for example, O(3)$_F$ instead of U(3)$_F$.
Then, the nonet fields in the present model will be replaced
with $({\bf 1}+{\bf 5})_S +{\bf 3}_A$ of O(3)$_F$.\footnote{
Note added: Based on an O(3) flavor symmetry, an extended 
version \cite{O3} of the present model has recently proposed.
The essential structure of the O(3) model is similar to 
that in the present U(3) model, 
and the substantial formulations have already been given
in the present paper.
Since the VEV matrices $\langle Y_f\rangle$ are diagonalized
as $U_f^\dagger \langle Y_f\rangle U_f = \langle Y_f\rangle^D$
and $U_f^T \langle Y_f\rangle U_f = \langle Y_f\rangle^D$
in the U(3) and O(3) models, respectively, we can use
Tr$[\langle \Phi_u\rangle]={\rm Tr}[\langle \Phi_u\rangle^D]
\propto {\rm diag}(\sqrt{m_u}, \sqrt{m_c}, \sqrt{m_t})$ and
Tr$[\langle \Phi_u\rangle\langle \Phi_u\rangle]=
{\rm Tr}[\langle \Phi_u\rangle^D \langle \Phi_u\rangle^D]
\propto {\rm diag}(m_u, m_c, m_t)$ in the U(3) model, while
we cannot use such relations in the O(3) model because
$U_f U_f^T \neq {\bf 1}$ in general.
Therefore, the U(3) model still has a considerable advantage
compared with the O(3) model.
}\   
How to extend the present model to a GUT model is also our 
future task.

\vspace{6mm}

\centerline{\large\bf Acknowledgments} 
The author would like to thank J.~Sato and T.~Yamashita for 
helpful discussions.
This work is supported by the Grant-in-Aid for
Scientific Research, Ministry of Education, Science and 
Culture, Japan (No.18540284).





\vspace{4mm}
\begin{thebibliography}{99}
%
%
\bibitem{tribi} 
P.~F.~Harrison, D.~H.~Perkins and W.~G.~Scott,
 Phys.~Lett. {\bf B458}, 79 (1999);
 Phys.~Lett. {\bf B530}, 167 (2002);
Z.~Z.~Xing, Phys.~Lett. {\bf B533}, 85 (2002);
P.~F.~Harrison and W.~G.~Scott,  Phys.~Lett. {\bf B535}, 163 (2003);
Phys.~Lett. {\bf B557}, 76 (2003);
E.~Ma, Phys.~Rev.~Lett. {\bf 90}, 221802 (2003);
C.~I.~Low and R.~R.~Volkas, Phys.~Rev. {\bf D68}, 033007 (2003);
X.-G.~He and A.~Zee,  Phys.~Lett. {\bf B560}, 87 (2003).
%
%
\bibitem{Koide90} Y.~Koide, Mod.~Phys.~Lett. {\bf A5}, 2319 (1990).
%
\bibitem{YK-Tanimoto96} Y.~Koide and M.~Tanimoto, {Z.~Phys. C}
{\bf 72}, 333 (1996).
%
%
\bibitem{YK-JHEP07}
Y.~Koide, JHEP {\bf 08}, 086 (2007). 
%
\bibitem{nonetVEV} Y.~Koide, Phys.~Lett. {\bf B662 } (2008) 43.
%
\bibitem{Koidemass} Y.~Koide, Lett.~Nuovo Cimento {\bf 34} (1982)  201;
Phys.~Lett. {\bf B120} (1983) 161;
Phys.~Rev. {\bf D28} (1983) 252.
%
\bibitem{FK-qmass} H.~Fusaoka and Y.~Koide, {Phys. Rev. D}
{\bf 57}, 3986 (1998).
%
\bibitem{PDG06} Particle Data Group, {J.~Phys.~G} {\bf 33}, 1 (2006) .
%
\bibitem{evol}
T.~P.~Cheng, E.~Eichten, and L.~F.~Li, Phys.~Rev. {\bf D9}, 2259
(1974);
M.~Marchacek and M.~Vaughn, Nucl.~Phys. {\bf B236}, 221 (1984);
M.~Olechowski and S.~Pokorski, Phys.~Lett. {bf B257}, 388 (1991);
H.~Arason {\it et al.}, Phys.~Rev. {\bf D46}, 3945 (1992);
V.~Barger, M.~S.~Berger, and P.~Ohmann, Phys.~Rev. {\bf D47}, 1093 (1993).
%
\bibitem{KamLAND} S.~Abe, {\it et al.}, KamLAND collaboration,
arXive:0801.4589.
%
\bibitem{MINOS} D.~G.~Michael {\it et al.}, MINOS collaboration,
Phys.~Rev.~Lett. {\bf 97}, 191801 (2006);
J.~Hosaka, {\it et al.}, Super-Kamiokande collaboration, Phys.~Rev. 
{\bf D74}, 032002 (2006).
%
\bibitem{sum_mnu} K.~Ichiki and Y.~Y.~Keum, JHEP {\bf 06}, 058 (2008).
%
\bibitem{O3} Y.~Koide, Phys.Lett. {\bf B665}, 227 (2008).
%
\end{thebibliography}
\end{document}